\documentclass{jfm}
\usepackage{graphicx}
\usepackage{epstopdf, epsfig}
\usepackage{xcolor}
\usepackage{array}
\usepackage{amsmath}
\usepackage{adjustbox}
\usepackage{mathrsfs}
\usepackage{bm}
\usepackage{natbib}
\usepackage{placeins}
\usepackage{hyperref}
\hypersetup{
  colorlinks=true,
  linkcolor=blue,
  citecolor=blue,
  filecolor=magenta,
  urlcolor=cyan
}

\shorttitle{Off-center gravity induces large-scale flow patterns in spherical RB}
\shortauthor{G. Wang, L. Santelli, R. Verzicco, D. Lohse, and R. J. A. M. Stevens}

\title{Off-center gravity induces large-scale flow patterns in spherical Rayleigh-B\'enard}
\author{Guiquan Wang\aff{1}
\corresp{\email{g.wang@utwente.nl}},
 Luca Santelli \aff{3},
 Roberto Verzicco\aff{1,2,3},
 Detlef Lohse\aff{1,4},
 \and Richard J. A. M. Stevens\aff{1}
 \corresp{\email{r.j.a.m.stevens@utwente.nl}} }

\affiliation{\aff{1}Physics of Fluids Group and Twente Max Planck Center, Department of Science and Technology, Mesa+ Institute, and J. M. Burgers Center for Fluid Dynamics, University of Twente, P.O. Box 217, 7500 AE Enschede, The Netherlands
\aff{2}Dipartimento di Ingegneria Industriale, University of Rome' Tor Vergata', Via del Politecnico 1, 00133 Rome, Italy
\aff{3}Gran Sasso Science Institute, Viale F. Crispi 7, 67100 L'Aquila, Italy
\aff{4}Max Planck Institute for Dynamics and Self-Organization, Am Fassberg 17, 37077 G\"ottingen, Germany}

\begin{document}

\maketitle

\begin{abstract}
Inspired by the hemispherical asymmetry observed in the Earth's inner core, we perform direct numerical simulations to study the effect of the gravity center offset in spherical Rayleigh-B\'enard convection. We find that even a minimal shift of the gravity center has a pronounced influence on the flow structures. When the gravity center is shifted towards the South, the co-latitudinal buoyancy component creates an energetic jet on the Northern side of the inner sphere that is directed towards the outer sphere. As a result, a large-scale meridional circulation is formed. However, surprisingly, the global heat flux is not affected by the changes in the large-scale flow organization induced by the gravity center offset. Our results suggest that the hemispherical core asymmetry is key to model the flow phenomena in the Earth's outer core and mantle.
\end{abstract}


\section{Introduction}\label{sec:introduction}
Seismological evidence shows an East-West asymmetry of the Earth's inner core \citep{tanaka1997degree,ohtaki2018seismological,deuss2014heterogeneity}. The hemispherical asymmetry is a prominent feature of Earth's inner core \citep{Monnereau1014, alboussiere2010melting, gubbins2011melting}. However, the effect of such an off-center gravity location on the flow structures in turbulent thermal convection has not been studied previously.
%\cite{gubbins2011melting} demonstrated that the mantle extracts heat from Earth's inner core at a spatially non-uniform rate. Such non-uniform heat flux patterns can affect plate tectonics, the creation of new plate margins, and volcanic activities \citep{schubert2001mantle}. 

We use spherical Rayleigh-B{\'e}nard (RB) as an idealized model system. The system consists of a fluid layer enclosed between two spherical shells, heated from the inner sphere and cooled from the outer one. Convection in spherical shells differs from the classical RB convection \citep{Ahlers2009, lohse2010small, Chilla2012} due to the curvature of the boundaries, non-uniform gravity, and the geometrical asymmetry between the boundary layer at the inner and outer spheres \citep{busse1970differential, spiegel1971convection,jarvis1995effects, tilgner1996high, shahnas2008convection, deschamps2010temperature, o2013comparison, Gastine2015}. 

Most studies on spherical RB convection focus on the case in which the gravity center coincides with the geometric one. \cite{busse1975patterns} used perturbation analysis to show the qualitative difference between convective flow patterns of odd and even spherical harmonic order just above the onset of convection. Subsequently, \cite{bercovici1989three} showed that these convective patterns persist for $Ra$ up to 100 times larger than the critical value. \cite{iwase1997interpretation} showed that the axisymmetric convective patterns break down for Rayleigh number $Ra>10^5$ when the flow starts to show time-dependent behavior. In this regime, sheet-like thermal plumes are formed near the inner and outer spheres, and these plumes undergo morphological changes into mushroom-like plumes when they eject from the boundary layers to the bulk \citep{bercovici1989three, yanagisawa2005rayleigh, Futterer2013, Gastine2015}. These plume dynamics are similar to the situation for classic RB convection \citep{puthenveettil_arakeri_2005, shishkina2008analysis,zhou2010physical, Chilla2012}. However, due to the asymmetries between the hot and cold surfaces in spherical shells, the mushroom-like plumes emitted from the outer sphere are thicker than those emitted from the inner sphere \citep{Gastine2015}. Nevertheless, studies have shown that there are similarities between spherical and classical planar RB convection. For example, \cite{Gastine2015} showed that the effective heat transport scaling exponent $\alpha$ in $Nu \sim Ra^\alpha$ in spherical RB has a similar $Ra$ number dependence as in planar RB, where \textcolor{black}{for $Ra\lesssim10^{11}$ typically} $0.28 \lesssim \alpha \lesssim 0.31$ \textcolor{black}{\citep{Ahlers2009}}.

However, the hemispherical asymmetry effect has so far been overlooked. To overcome this shortcoming, in this study, we use three-dimensional direct numerical simulations to investigate the effect of the gravity center offset on the flow structures and heat transfer in spherical RB convection, \textcolor{black}{and we will indeed find a very strong effect, which we can then theoretically explain. The paper is organized as follows: In section \ref{sec:Numerical}, we explain the numerical method and motivate what parameters we chose for the study. In section \ref{subsec:jet_generation} and \ref{subsec:Spectral analysis}, we show the large-scale flow pattern induced by the off-center gravity. The effect of the large-scale structure on heat transfer is shown in section \ref{subsec:Nu_Tprofile}. In section \ref{subsec:gravity difference}, we study the effect of $n_g$ and $Ra$. The paper ends with conclusion and an outlook.}

\section{Numerical method and parameters}\label{sec:Numerical}

\subsection{Setup of convection in spherical shells} \label{sec:set-up}
\begin{figure}
\centering
\includegraphics[width=7.5cm, trim={0cm 0cm 0cm 0cm},clip]{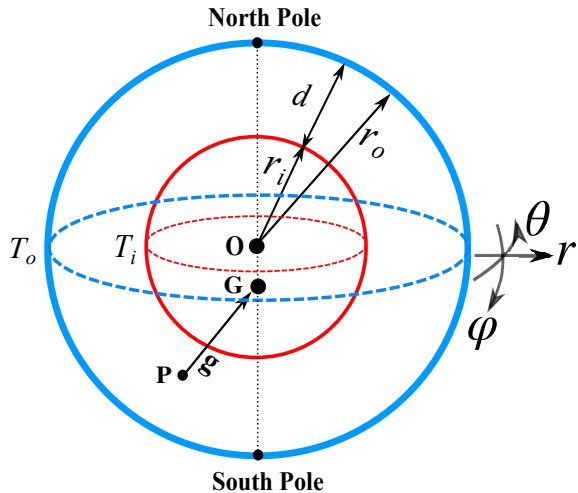}
\caption{Schematic of spherical RB convection in between two concentric spheres with radius ratio $\eta=r_i/r_o$ between the inner and outer sphere and gap size $d=r_o-r_i$. A no-slip boundary condition with constant temperature is used on the inner (hot) and outer (cold) sphere. In spherical coordinates the longitudinal, co-latitudinal, and radial directions are represented by $\hat{\theta}$, $\hat{\varphi}$ and $\hat{r}$, respectively. The gravity center $G$ is offset from the geometric center $ O $ and $P$ is an arbitrary fluid point in the spherical shell.}
\label{fig:sphere_coordinate}
\end{figure}
The spherical RB geometry is schematically illustrated in figure \ref{fig:sphere_coordinate}. Fluid fills a spherical shell between the inner sphere of radius $r_i$ and the outer sphere of radius $r_o$. The radius ratio between the inner and the outer sphere is given by 
\begin{align}
\label{eq:eta}
\eta=r_i/r_o.
\end{align}
In this study we consider a fixed aspect ratio $\eta=0.3$, which is representative for Earth's outer core. The surface temperature of the inner and outer spheres is kept constant at $T_i$, and $T_o$, respectively, with $T_i > T_o$. No-slip boundary conditions are imposed on both boundaries. Point $G$ indicates the gravity center, which is offset from the geometric center $O$. The gravity shift is defined as:
\begin{align}
\varepsilon=|\mathbf{OG}|/r_i,
\label{eq:epsi}
\end{align}
where $|\mathbf{OG}|$ indicates the displacement of the gravity compared to the geometrical center, see figure \ref{fig:sphere_coordinate}.

The dynamics of spherical RB convection is controlled by the Rayleigh and the Prandtl numbers
\begin{equation}
Ra=\frac{\beta g_o \Delta T d^3}{\kappa \nu},~ Pr=\frac{\nu}{\kappa},
\label{eq:Ra_Nu}
\end{equation}
where $\beta$ is the thermal expansion coefficient, $g_o$ is the surface averaged gravity at the outer sphere, $\nu$ is the kinematic viscosity, and $\kappa$ is the thermal diffusivity. We normalize the \textcolor{black}{fields and distances} by the length scale $d=r_o-r_i$, the temperature difference $\Delta T$ between the inner and outer sphere, and the free-fall velocity $U=\sqrt{\beta g_o \Delta T d}$.

\subsection{\textcolor{black}{Underlying dynamical equation and numerical method}} \label{sec:Numerical_method}
We solve the Navier-Stokes equations under the Boussinesq approximation in spherical coordinates, which in dimensionless form read:
\begin{equation}
  \frac{\partial \mathbf{u}}{\partial t}+\mathbf{u} \cdot \nabla \mathbf{u}=-\nabla p+\sqrt{\frac{P r}{R a}} \nabla^{2} \mathbf{u}+ \mathbf{C_{g}} g T ~~,~~\nabla \cdot \mathbf{u}=0, \\
  \label{eq:NS_2}
\end{equation}
\begin{equation}
  \frac{\partial T}{\partial t}+\mathbf{u} \cdot \nabla T=\frac{1}{\sqrt{{Ra Pr}}} \nabla^{2} T.
  \label{eq:NS_3}
\end{equation}
where $\mathbf{u}$, p, $T$, and $\mathbf{C_{g}}g$ denote the fluid velocity, pressure, temperature, and gravitational acceleration acting in the three directions. The coefficients $C_{g,i}$ ($i=1,2,3$ correspond to the $\hat{\theta}$, $\hat{r}$, and $\hat{\varphi}$ directions respectively) denote the decomposition of $g$, which is detailed in appendix \ref{appB}. We consider the following gravity profile
\begin{equation}
  g=(L_{GP}/r_o)^{n_g},
  \label{eq:n_g}
\end{equation}
in which $L_{GP}=|\mathbf{GP}|$ indicates the distance between a fluid point and the geometrical center, see figure \ref{fig:sphere_coordinate}, and we use \textcolor{black}{$n_g=-2$, $-1$, $0$, and $1$}. The value of $n_g$ can be inferred by physical considerations on the mass ratio between the nucleus ($r<r_i$) and the spherical shell ($r_i<r<r_o$). The gravity is constant ($n_g=0$) when the mass of the spherical shell is negligible. This assumption is typically used to model the Earth's mantle \citep{bercovici1989three}. If the mass is centrally condensed, one obtains $n_g=-2$. \cite{Christensen2006} showed that, only for this gravity distribution function, there is an analytical relation between the viscous dissipation rate and the Nusselt number ($Nu$). When the density is constant within the fluid layer, and no mass is contained in the inner sphere, gravity is directly proportional to the radial coordinate $r$ ($n_g=1$) \citep{tilgner1996high}. For the sake of completeness, $n_g=-1$ is also included.

\begin{figure}
\centering
\includegraphics[width=12cm]{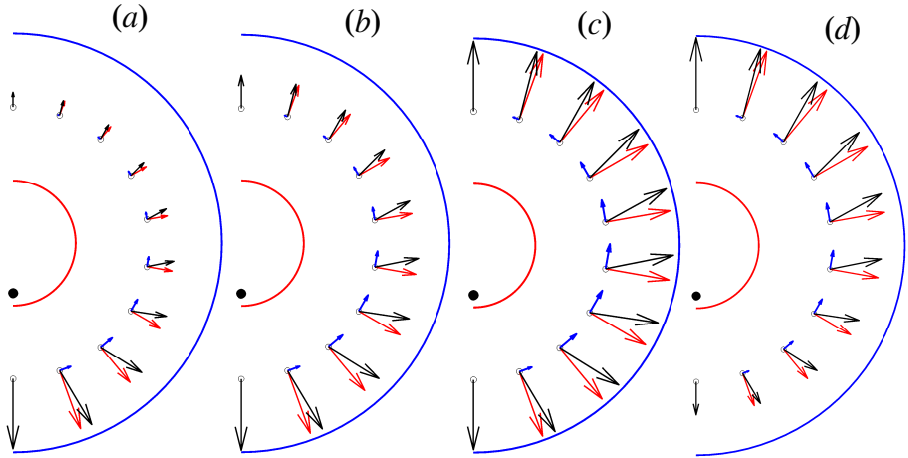}
 \caption{The radial ($C_{g,r}g \hat{r}$, red arrow) and co-latitudinal ($C_{g,\varphi}g \hat{\varphi}$, blue arrow) decomposition of buoyancy $g \mathbf{GP}$ (black arrow) in Eq.\ \eqref{eq:buoyancy_components} for $\varepsilon=0.8$. Panels (a-d) indicate the buoyancy components on the mid-plane of the spherical shell for (a) $n_g=-2$, (a) $n_g=-1$, (a) $n_g=0$, and (d) $n_g=1$, see Eq.\ \eqref{eq:n_g}. The length of the arrows is scaled by the magnitude of $g$ in the midpane.}
\label{fig:gravity_sketch}
\end{figure}
For convenience, we always shift the gravity center $G$ to the South. In that case, the three components of buoyancy coefficient can be written as
\begin{equation}
  C_{g,i}g=(0,~r-r_i \varepsilon \cos{\varphi},~-\varepsilon{r_i}\sin\varphi)r_o^{-n_g}L_{GP}^{n_g-1}.
  \label{eq:buoyancy_components}
\end{equation}
where $\varepsilon$ (Eq.\ \ref{eq:epsi}) indicates the gravity center offset. Figure \ref{fig:gravity_sketch} shows the radial ($C_{g,r}g \hat{r}$) and co-latitudinal ($C_{g,\varphi}g \hat{\varphi}$) buoyancy coefficients for $\varepsilon=0.8$ on the mid-plane of the spherical shell for different $n_g$. It shows that $g$ monotonically increases from the North to the South Pole when $n_g<0$ (panels (a,b)). For $n_g>0$ (panel (d)) the opposite trend is observed, while for $n_g=0$ the gravity is constant as function of the  co-latitude location. %{\color{red}short description off-center gravity}

The governing equations (\ref{eq:NS_2}--\ref{eq:NS_3}) are discretized by a staggered central second-order finite-difference scheme in spherical coordinates. The numerical scheme is based on the method by \cite{verzicco1996finite}, which has recently been extended to spherical coordinates by \cite{santelli2020finitedifference}. The advantage of this scheme is that it allows arbitrary non-uniform grids in the radial and co-latitudinal directions. Here it is worthwhile to mention that the singularities at the poles are prevented by introducing a new set of quantities $(u_\theta,u_r r^2, u_\varphi \sin\varphi)$, which results in trivial boundary conditions $u_\varphi \sin\varphi=0$ at the North ($\varphi=0$) and South Poles ($\varphi=\pi$) in the co-latitudinal direction. The code has been validated carefully by \cite{santelli2020finitedifference}, and we refer to that work for details on the method. For additional validation results relevant to the flows considered here, we refer the reader to Appendix \ref{appA}.

\subsection{\textcolor{black}{Choice of parameters}}

\begin{table}
\centering
\vspace{0.1cm}

\begin{adjustbox}{width=0.9\textwidth}
\begin{tabular}{cccccccc}
$\quad Ra \quad$     & $\quad n_g \quad$ & $\quad N_\theta \times N_r \times N_\varphi \quad$ & $\quad \varepsilon \quad$ & $\quad Nu \quad$  & $\quad \langle {Nu_h} \rangle_r \quad $ & $\quad Re_{rms} \quad$ \\
\hline 
       &    &             &        &    &    &    \\
$1\times10^7$ & $-2$  & $501\times101\times251$  & 0       & 16.09 & 16.05 & 919.5 \\
       &    &             & 0.2      & 16.18 & 16.17 & 965.7 \\
       &    &             & 0.4      & 16.23 & 16.22 & 976.9 \\
       &    &             & 0.8      & 16.20 & 16.08 & 936.8 \\
       &    &             &        &    &    &    \\
       
$1\times10^7$ & $-1$  & $401\times91\times201$  & 0       & 11.89 & 11.90 & 657.5 \\
       &    &             & 0.2      & 11.91 & 11.89 & 692.5 \\
       &    &             & 0.4      & 11.96 & 11.95 & 702.6 \\
       &    &             & 0.8      & 11.41 & 11.14 & 526.6 \\
       &    &             &        &    &    &    \\
       
$1\times10^7$ & $0$  & $301\times81\times151$  & 0       & 8.73 & 8.77 & 485.6 \\
       &    &             & 0.2      & 8.91 & 8.91 & 497.6 \\
       &    &             & 0.4      & 8.97 & 8.96 & 494.4 \\
       &    &             & 0.8      & 8.81 & 8.74 & 409.3 \\
       &    &             &        &    &    &    \\

$1\times10^7$ & $1$  & $321\times85\times161$  & 0       & 7.04 & 7.05 & 358.4 \\
       &    &             & 0.2      & 6.93 & 6.89 & 356.3 \\
       &    &             & 0.4      & 7.06 & 7.03 & 359.5 \\
       &    &             & 0.8      & 7.10 & 7.06 & 319.4 \\
       &    &             &        &    &    &    \\
       
$1\times10^8$ & $0$  & $501\times101\times251$  & 0       & 17.36 & 17.38 & 1487.3 \\
       &    &             & 0.05     & 17.37 & 17.42 & 1501.1 \\
       &    &             & 0.1      & 17.30 & 17.29 & 1505.2 \\
       &    &             & 0.2      & 17.15 & 17.15 & 1529.0 \\
       &    &             & 0.4      & 17.31 & 17.24 & 1513.7 \\
       &    &             & 0.8      & 16.62 & 16.53 & 1302.8 \\
\end{tabular}
\end{adjustbox}

 \caption{Details of the simulations. The columns from left to right indicate: $Ra$, the gravity profile exponent $n_g$ (see Eq.\ \eqref{eq:n_g}), the number of grid points in the longitudinal, radial, and co-latitudinal direction $N_{\theta} \times N_r \times N_\phi$, the shift of the gravity center with respect to the geometrical center $\varepsilon$ (see Eq.\ \eqref{eq:eta}), the average heat transfer across the inner and outer sphere ($Nu$, see Eq.\ \eqref{eq:Nu}), $\langle Nu_h \rangle$ (see Eq.\ \eqref{eq:Nu_h}), and $Re_{rms}$ (see Eq.\ \eqref{eq:Re_rms}).}
 \label{tab:Table_3} 
 
\end{table}

To study the effect of the gravity center location on the flow dynamics in spherical RB convection, we considered different gravity distributions ($n_g \in \{-2,-1,0,1\}$) in Eq.\ \eqref{eq:NS_2}  and Eq.\ \eqref{eq:n_g}) and gravity center offset ($\varepsilon$ from $0$ to $0.8$, see Eq.\ \eqref{eq:epsi}) for $Ra=10^7$ and $10^8$. All simulations in this study are for $Pr=1$. \textcolor{black}{We realize that inside the outer core or the mantle $Pr$ in fact is quite different, but the objective of this study is not to have a one-to-one comparison with a geophysical situation, but to make out the physics of the off-center gravity.}

To ensure that the flow is fully resolved, we place a sufficient number of computational grid points in the bulk and the boundary layers \citep{stevens2010radial}. As we will use spectral analysis to study the flow structures, we use a uniform grid in the longitudinal and co-latitudinal directions. In the radial direction, the grid cells are clustered towards the inner and outer sphere to ensure the boundary layers are adequately resolved \citep{shishkina2010boundary}.

We calculate the $Nu$ number from the normalized averaged temperature gradients at the inner and outer sphere as follows \citep{Gastine2015}:
\begin{equation}
  Nu_i = \left. -\eta \frac{d \langle T \rangle_s}{dr} \right| _{r_i},~Nu_o=\left. -\frac{1}{\eta} \frac{d \langle T \rangle_s}{dr} \right| _{r_o}.
  \label{eq:Nu}
\end{equation}
Here $\langle ... \rangle_s$ represents the average over a spherical surface. The difference in the $Nu$ number obtained at the inner and outer sphere is always less than $0.2\%$. Besides, we verify that the $Nu$ number calculated at the spheres is the same as the value obtained from $Nu_h(r)$
\begin{equation}
  Nu_h(r)=\frac{\langle u_rT \rangle_s-\kappa \partial_r \langle T \rangle_s}{-\kappa \partial_r T_c},
  \label{eq:Nu_h}
\end{equation}
where $T_c(r)=\eta/[(1-\eta)^2r]-\eta/(1-\eta)$ is the \textcolor{black}{purely} conductive temperature profile for constant temperature boundary conditions. The volume-averaged root mean square (rms) Reynolds number is given by
\begin{equation}
  Re_{rms} = (Ra/Pr)^{0.5} \overline{\sqrt{\langle u_i u_i \rangle}},~~~ i=1,2,3.
  \label{eq:Re_rms}
\end{equation}
The details of the simulations considered in this study are summarized in table \ref{tab:Table_3}.

%\section{Results}\label{sec:Results}
\section{Generation of convective jet and large-scale circulation}\label{subsec:jet_generation}

As discussed in section \ref{sec:Numerical_method} above the radial and co-latitudinal buoyancy coefficients depend on the gravity center offset $\varepsilon$, see Eq.\ (\ref{eq:epsi}) and the gravity distribution $n_g$, see Eq.\ (\ref{eq:n_g}). To disentangle these effects, we first examine the effect of the gravity center offset for constant gravity ($n_g=0$) at $Ra=10^8$ in sections \ref{subsec:jet_generation}, \ref{subsec:Spectral analysis} and \ref{subsec:Nu_Tprofile}. The effect of $n_g$ and $Ra$ is studied in section \ref{subsec:gravity difference}.

%\subsection{Generation of convective jet and large-scale circulation}\label{subsec:jet_generation}
%
\begin{figure}
\centering
\includegraphics[width=12cm]{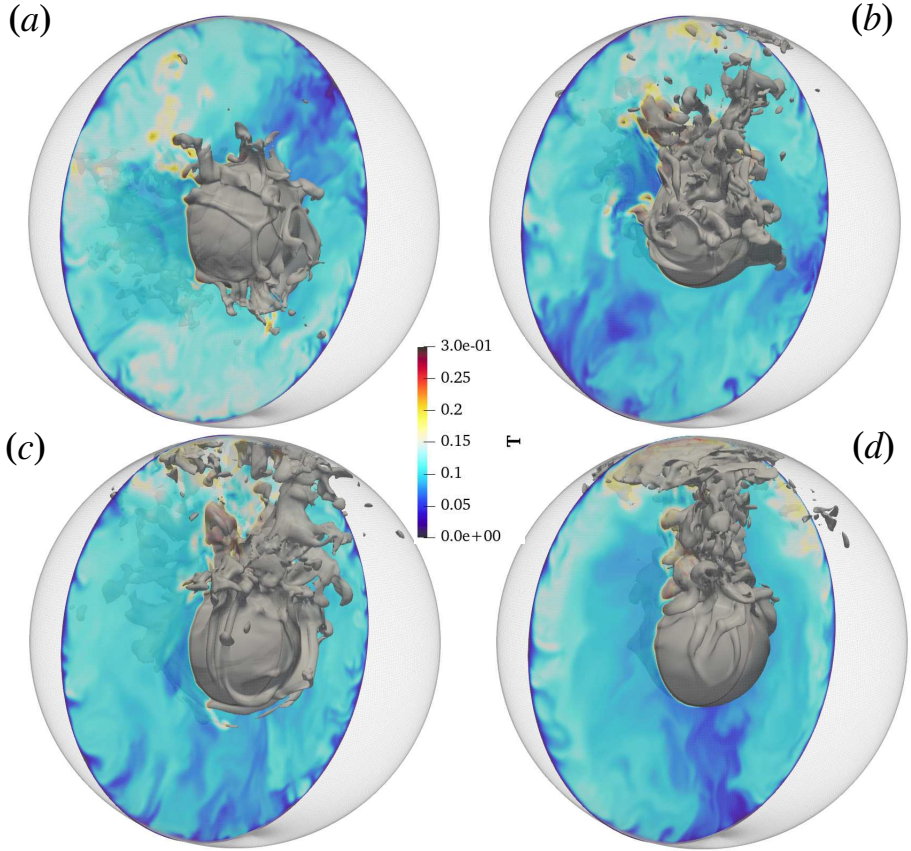}
 \caption{The temperature isosurface for $T=0.2$ is indicated in grey, while the color indicates the temperature field in the meridian cut for $Ra= 10^8$, $n_g=0$, and (a) $\varepsilon= 0$, (b) $\varepsilon= 0.05$, (c) $\varepsilon=0.4$, and $\varepsilon=0.8$, respectively.}
\label{fig:contour_T_jet}
\end{figure}
\begin{figure}
\centering
\includegraphics[width=12cm]{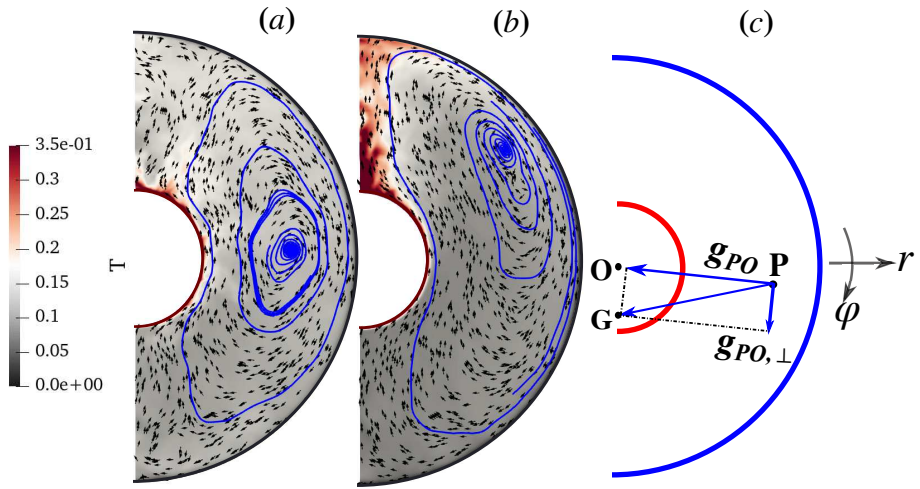}
 \caption{The vector field (black arrows) and streamlines (blue curves) of the average velocity in the meridian plane for $Ra=10^8$, $n_g=0$, and (a) $\varepsilon=0.05$ and (b) $\varepsilon=0.8$; (c) Schematic illustration of gravity decomposition $\mathbf{g}=\mathbf{g_{PO}}+\mathbf{g_{PO,\perp}}$ when the gravity center $G$ does not coincide with the geometrical center $O$.}
\label{fig:LSW_jet}
\end{figure}

Figure \ref{fig:contour_T_jet}(a) shows that uniformly distributed long and thin sheet-like thermal structures are generated at the inner sphere when the gravity center coincides with the geometrical center ($\varepsilon=0$). However, figure \ref{fig:contour_T_jet}(b) reveals that a small offset of the gravity center ($\varepsilon=0.05$) is sufficient to induce an asymmetric flow pattern in which hot plumes are preferentially emitted from the North Pole region of the inner sphere. The jet ejects hot plumes from the inner to the outer sphere and increases in strength when the gravity center offset, i.e.\ $\varepsilon$, is increased. Figure \ref{fig:LSW_jet} shows that due to ascending thermal plumes along the inner sphere and descending thermal plumes along the outer sphere, a large-scale circulation is formed. Furthermore, the figure shows that for larger $\varepsilon$ the flow in the Southern hemisphere becomes thermally stratified, due to which the number of plumes in this region is reduced.

To understand the generation of the convective jet and corresponding large-scale circulation, we decompose the gravity vector $\mathbf{g}$ into $\mathbf{g_{PO}}$ and $\mathbf{g_{PO,\perp}}$
\begin{equation}
  \mathbf{g}=\mathbf{g_{PO}}+\mathbf{g_{PO,\perp}}
\label{eq:norm_2}
\end{equation}
where the gravity component $\mathbf{g_{PO}}$ points towards the geometric center $\mathbf{O}$, while the other component $\mathbf{g_{PO,\perp}}$ works in the co-latitudinal direction, see figure \ref{fig:LSW_jet}(c). When the gravity center $\mathbf{G}$ coincides with the geometrical counterpart $\mathbf{O}$ ($\varepsilon=0$), $|\mathbf{g_{PO}}|$ is uniformly distributed in the co-latitudinal direction and $\mathbf{g_{PO,\perp}} = 0$. However, as shown in figure \ref{fig:gravity_sketch}, a co-latitudinal gravity component is introduced when the gravity center $\mathbf{G}$ is displaced Southwards. Figure \ref{fig:gravity_sketch}(c) shows that this co-latitudinal component is strong in the equatorial region. This gravity component drives the ascending hot thermal plumes towards the North Pole of the inner sphere where the convective jet is formed, see figure \ref{fig:LSW_jet}(b). Once the hot convective jet is impinging on the outer sphere, the flow descends along the outer sphere, thereby creating a strong clockwise meridional circulation as shown in figure \ref{fig:LSW_jet}(a,b).

\section{Modal analysis to identify large-scale structures}\label{subsec:Spectral analysis}

\begin{figure}
\centering
\includegraphics[width=13.6cm]{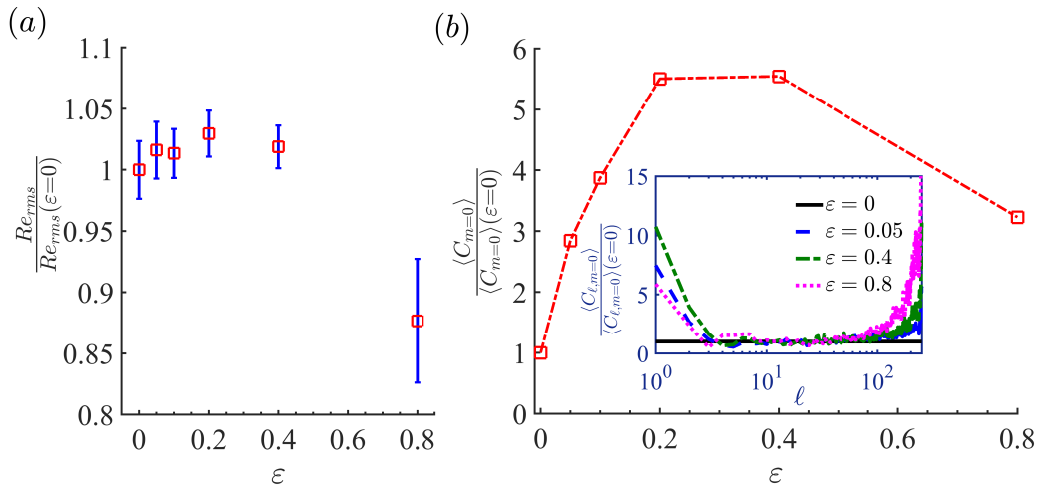}
 \caption{(a) Volume averaged $Re_{rms}(\varepsilon)/Re_{rms}(\varepsilon=0)$ as a function of $\varepsilon$ for $Ra=10^8$ and $n_g=0$. The error bars indicate the corresponding standard deviations. (b)$\langle C_{m=0} \rangle$ of $\mathrm{TKE}$, normalized by $\langle C_{m=0} \rangle$ for $\varepsilon=0$. The inset shows $\langle C_{\ell,m=0} \rangle$, normalized by $\langle C_{\ell,m=0} \rangle$ for $\varepsilon=0$, as a function of the spherical harmonic order $l$, for different $\varepsilon$.}
\label{fig:SPH_lm}
\end{figure}

When the gravity center is moved towards the South Pole (increasing $\varepsilon$), the total turbulent kinetic energy ($\mathrm{TKE}$), which is defined as
\begin{equation*}
  \mathrm{TKE} = \frac{Ra}{Pr} \frac{1}{2} \overline{\langle u_i u_i \rangle},~~~ i=1,2,3
  \label{eq:TKE}
\end{equation*}
first increases slightly ($\varepsilon \leq 0.4$), before it decreases considerably  ($\varepsilon = 0.8$, see figure \ref{fig:SPH_lm}(a). To study the effect of the off-center gravity on the distribution of the $\mathrm{TKE}$, we perform a modal analysis. In spherical coordinates, a surface spherical harmonic function is the expansion of plane waves, where the order $m$ and degree $\ell$ represent the wavenumbers along with the longitudinal and co-latitudinal directions, respectively. $\langle C_{\ell,m} \rangle$ represents the radial integration and temporally averaged power spectrum, which is detailed in the appendix \ref{appC}.

Figure \ref{fig:SPH_lm}(a) shows that the average $\mathrm{TKE}$ is almost unchanged for $\varepsilon \leq 0.2$. However, the distribution among the different modes is changed. As we have seen in figure \ref{fig:LSW_jet}, a large-scale circulation is formed when $\varepsilon>0$. Figure \ref{fig:SPH_lm}(b) shows that the $m=0$ mode in the longitudinal direction, which indicates the large-scale mode, strongly increases with $\varepsilon$. For $\varepsilon > 0.4$ both the $m=0$ mode and the total $\mathrm{TKE}$ decrease sharply due to the thermal stratification that is formed in the Southern hemisphere. Interestingly, only a small gravity center offset is required to form the convective jet and the corresponding meridional circulation. The inset of figure \ref{fig:SPH_lm}(b) shows the distribution of the $m=0$ mode over the different co-latitudinal structures ($\langle C_{\ell, m=0} \rangle$), which shows that the large-scale circulation is dominated by the co-latitudinal modes with $\ell \leq 3$. With increasing $\varepsilon$, $\langle C_{\ell, m=0} \rangle$ is enhanced for low ($\ell \leq 3$) and high wavenumbers ($\ell \geq 70$). 

%%%%%%%%%%%%%%%%%%%%%%%%%%%%%%%%%%%%%%%%%%%%%%%%%%%%%%%%%%
\section{Effect of large-scale flow on heat transfer distribution} \label{subsec:Nu_Tprofile}

\begin{figure}
\centering
    \includegraphics[width=12cm]{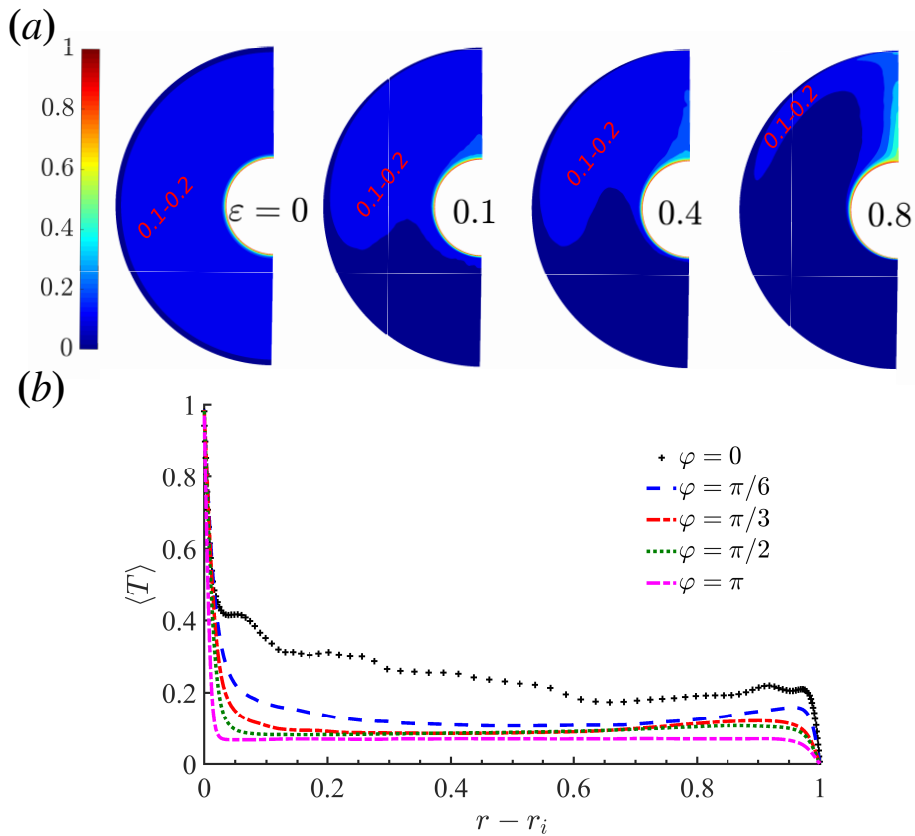}
 \caption{(a) Mean temperature $\langle T \rangle (r,\varphi)$ in the meridian plane for $Ra=10^8$, $n_g=0$, and $\varepsilon=$0, 0.1, 0.4, and 0.8. Panel (b) shows mean temperature profiles at different co-latitudinal locations for $\varepsilon=0.8$.}
\label{fig:T_profile}
\end{figure}

As we have shown above, the gravity center offset strongly influences the flow structures. This raises the question of how the local temperature profiles and heat fluxes are affected. Figure \ref{fig:T_profile}(a) shows the mean temperature in the meridian plane. For $\varepsilon=0$ the temperature is uniform in the co-latitude direction. However, with increasing $\varepsilon$, the effect of the convective jet, which brings hot fluid from the inner to the outer sphere, becomes more pronounced. The effect of the jet is more clearly visible in figure \ref{fig:T_profile}(b), which shows the temperature profiles at selected angular positions in the co-latitudinal direction. The figure shows that due to the jet and corresponding large-scale circulation temperature inversions in the Northern hemisphere near the outer sphere ($0 \le \varphi \le \pi/2$ within $0.6<r-r_i<0.95$) are formed. Besides, the temperature profiles indicate that the flow becomes thermally stratified in the bulk.

\begin{figure}
\centering
\includegraphics[width=12.5cm, trim={0cm 0cm 0cm 0cm},clip]{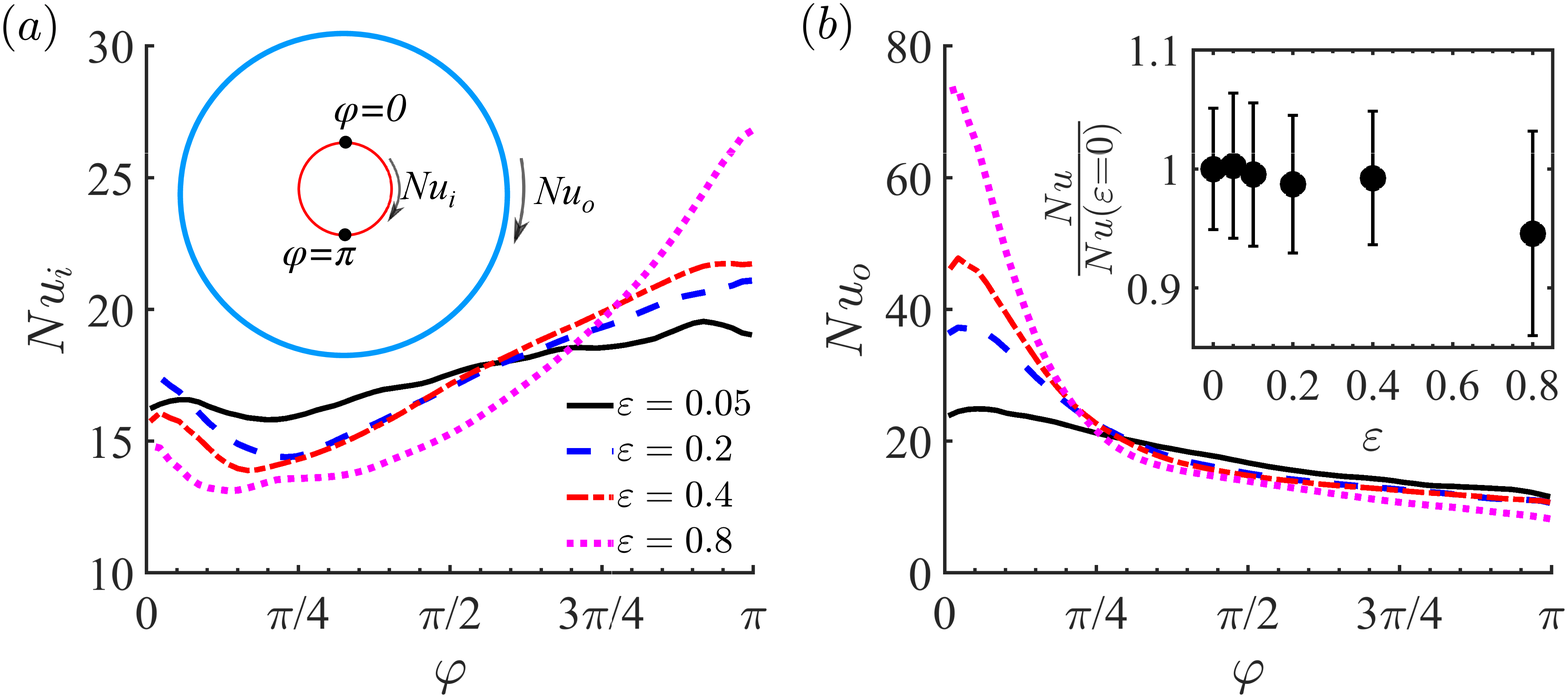}
 \caption{(a) $Nu$ number on the inner and (b) outer sphere as a function of the co-latitudinal direction for $Ra=10^8$ and $n_g=0$. The inset of (b) shows $Nu (\varepsilon) / Nu (\varepsilon=0)$ as a function of $\varepsilon$. The error bar represents the standard deviation. We note the error bars for standard errors are smaller than the size of symbols.}
\label{fig:Nu_io_Q}
\end{figure}

Next, we examine how the local heat fluxes depend on $\varepsilon$. Figure \ref{fig:Nu_io_Q} shows the Nusselt number along the inner ($Nu_i$) and outer spheres ($Nu_o$). It is observed that the local heat fluxes on the outer sphere decrease from $\varphi=0$ (North Pole) to $\varphi=\pi$ (South Pole), whereas the opposite trend, is observed on the inner sphere. This happens because the hot convective jet impinges on the outer boundary layer, resulting in a thinner thermal boundary layer and larger heat flux than the situation without a convective jet. Surprisingly, the formation of the large-scale circulation and corresponding non-uniform heat flux distribution have a limited influence on the overall hear flux, see inset figure \ref{fig:Nu_io_Q}(b), although some reduction in the heat transport is observed for $\varepsilon=0.8$ when the flow becomes more thermally stratified. 

\section{Effect of $Ra$ and gravity profile} \label{subsec:gravity difference}

So far, we focused on the effect of the gravity center offset on the flow structures. In this section, on the one hand, we demonstrate that the convective jet and large-scale flow structures are formed for a wide range of $n_g$ and $Ra$ and that the intensity of the large-wavelength structures is significantly modified by the control parameters. However, surprisingly, the global heat fluxes are insensitive to $\varepsilon$ for all considered $n_g$ and $Ra$.

\begin{figure}
\centering
\includegraphics[width=12cm]{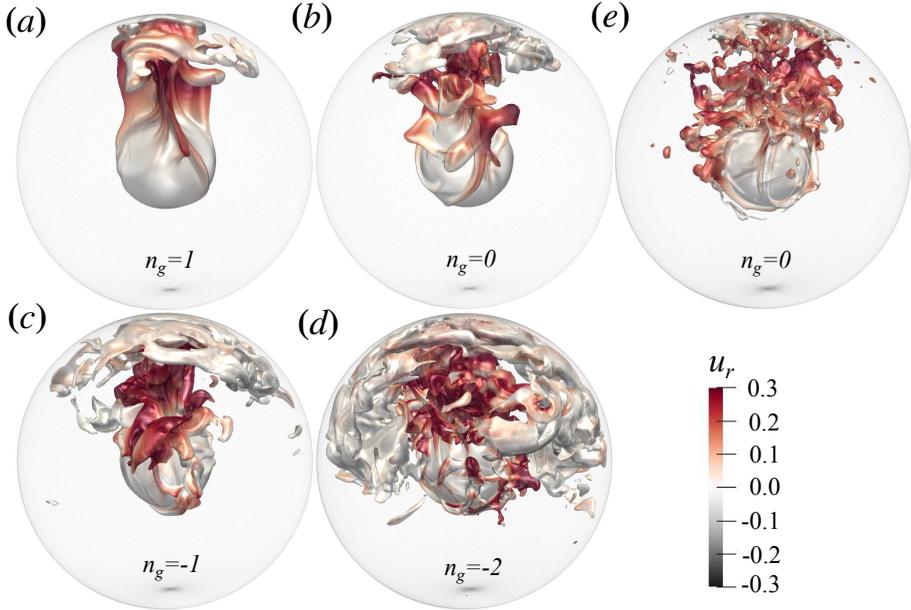}
 \caption{(a-d) Instantaneous isosurface of $T=0.2$ for different gravity distributions from $n_g=1$ to $n_g=-2$ (see Eq.\ \ref{eq:n_g}) for $Ra=10^7$ and $\varepsilon=0.4$. Panel (e) shows $Ra=10^8$, $n_g=0$, and $\varepsilon=0.4$ for comparison. The color indicates $u_r$ at the isosurface. (Two movies for $Ra=10^7$, $n_g=0$ and $\varepsilon=0.8$ are provided; see supplementary movie 1 and 2.)}
\label{fig:Gravity_ng_isocontour}
\end{figure}

Figures \ref{fig:Gravity_ng_isocontour}(a-d) show visualizations of the convective jet for $Ra=10^7$, $\varepsilon=0.4$, and different gravity profiles. The figure shows that the formation of the convective jet with $u_r>0$ in the North Pole region when the gravity center shifts to the South. The jet becomes more turbulent when $n_g$ is decreased. As previously shown in figure \ref{fig:gravity_sketch}, although the co-latitudinal buoyancy component $C_{g,\varphi}g \hat{\varphi}$ points to the North Pole in all circumstances, the intensity of the radial buoyancy coefficient $C_{g,r}g \hat{r}$ is strongly dependent on the co-latitude, e.g. $C_{g,r}g$ at the South Pole is the highest when $n_g=-2$. It is well accepted that the higher radial component tends to induce more intense thermal plumes. However, the convective jet is still observed in the North Pole region rather than the South Pole region when $n_g=-2$, as shown in figure \ref{fig:Gravity_ng_isocontour}(d). This confirms that the convective jet and large-scale circulation are a result of the co-latitudinal buoyancy component.
\begin{figure}
\centering
\includegraphics[width=13.5cm]{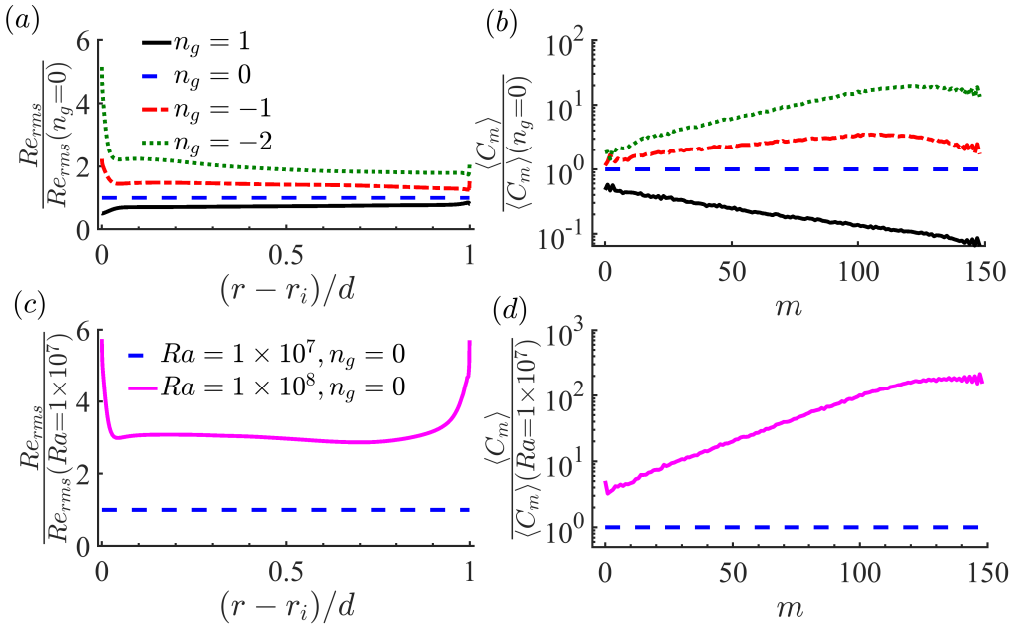}
 \caption{ (a,c) $Re_{rms}(r)$ as function of the gap and (b,d) $\mathrm{TKE}$ spectrum of $\langle C_{m} \rangle$ for $\varepsilon=0.4$. (a,b) $Ra=10^7$ and different $n_g$; (c,d) $n_g=0$ and different $Ra$. Figures are normalized by $Ra=10^7$ with $n_g=0$. }
\label{fig:Cm_ng}
\end{figure}

The corresponding $Re_{rms}(r)$ presented in figure \ref{fig:Cm_ng}(a) shows that the turbulent intensity is greatly enhanced with decreasing $n_g$. This reveals that the flow becomes more turbulent when $n_g$ is decreased. This is because the buoyancy coefficients between the inner and outer sphere differ more for lower $n_g$. Correspondingly, panel (b) shows that the $\mathrm{TKE}$ spectrum of $\langle C_{m} \rangle$ increases for all wavenumbers. We note that $\langle C_{m=0} \rangle$, which dominates the TKE and indicates the strength of the large-scale circulation, increases with decreasing $n_g$. With increasing $Ra$ from $10^7$ to $10^8$, the normalized $Re_{rms}(r)$ is enhanced as shown in figure \ref{fig:Cm_ng}(c). The enhancement happens for all wavenumbers, as shown in panel (d), especially for the high wavenumber structures. Thus, for a constant $\varepsilon$, the large-scale circulation is more energetic for smaller $n_g$ and higher $Ra$ due to the relatively higher co-latitudinal component of buoyancy force as illustrated in figure \ref{fig:LSW_jet}(c). 

\begin{figure}
\centering
\includegraphics[width=10cm]{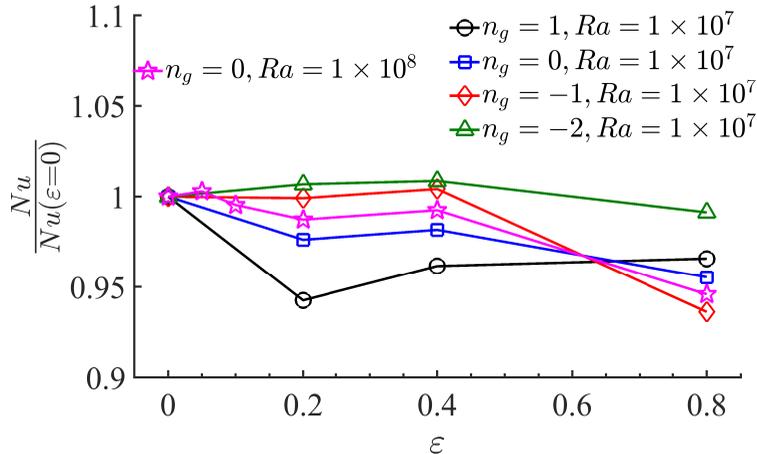}
 \caption{$Nu (\varepsilon) / Nu (\varepsilon=0)$ as a function of $\varepsilon$ for different $Ra$ and gravity profiles $n_g$, see Eq.\ \eqref{eq:n_g}.}
 \label{fig:Nu_epsilon}
\end{figure}

Although the buoyancy coefficient components are significantly different for various $n_g$ considered here, we find that the convective jet and large-scale structures are always formed for off-center gravity ($\varepsilon>0$). The generation of large-scale structures can induce highly non-uniform co-latitudinal heat flux distribution. We previously showed that the global heat transfer is relatively insensitive to $\varepsilon$ for $n_g=0$ and $Ra=10^8$. Figure \ref{fig:Nu_epsilon} shows that a similar trends are observed for different $n_g$ and $Ra$. In particular, for $\varepsilon \leq 0.8$ the difference between $Nu(\varepsilon>0)$ and $Nu(\varepsilon=0)$ is less than $8\%$, which indicates that the effect of the changed large-scale flow structures on the overall heat transport is relatively limited.

\section{\textcolor{black}{Summary, conclusions, and outlook}}\label{sec:Conclusion}
We demonstrated using direct numerical simulations that even a small shift of the gravity center location can introduce pronounced changes in the flow organization and local heat fluxes in spherical RB convection. When the gravity center is moved to the South, a strong convective jet on the Northern side of the inner sphere is formed, which leads to a large-scale flow organization with hot ascending fluid close to the inner sphere and cold descending fluid close to the outer sphere. The formation of these large-scale flow structures has been confirmed by modal analysis. Due to the large-scale flow organization, the heat transfer at the inner and outer sphere is highly non-uniform. In particular, around the North Pole, the heat flux is enhanced on the outer sphere due to the impingement of the convective jet. However, surprisingly, the global heat flux is relatively insensitive to the gravity center shift. These findings do not seem to depend much on $Ra$ (over our explored range) or the employed gravity profile.

Although spherical RB convection is a simplified representation of what may happen in the Earth, our study shows that even small hemispherical asymmetries may have important effects on the large-scale flow organization. In Earth's core such a hemispherical asymmetry is found as the crystallizing Western and melting Eastern hemispheres result in the East-West asymmetry. The mass center gradually translates to the West from its geometric center \citep{alboussiere2010melting, Monnereau1014}. Our study indicates that such a shift in the gravity center location can results in a high-temperature region on the Eastern side of the core and a low-temperature region on the Western side, which would enhance the asymmetry. In addition, the impingement of a convective jet with the core-mantle boundary can significantly enhance the local heat transfer from the inner core to the mantle, which can trigger volcanic eruptions on the Eastern side \citep{schubert2001mantle,gubbins2011melting}. Hence, this study should be considered a first step towards realizing a numerical model aimed at capturing the dynamics in the Earth's outer core and mantle due to the hemispherical asymmetry, even though many additional features, such as rapid rotation \citep{julien2012heat,aurnou2015rotating, Gastine2016, long_mound_davies_tobias_2020}, Earth's dynamo \citep{Aubert2017}, the effect of the magnetic field \citep{zhang2000magnetohydrodynamics}, and temperature-dependent \citep{tackley1996effects,zhong2000role,calkins2012influence} fluid properties are not yet included.

\FloatBarrier
\section*{Acknowledgements}\label{sec:Acknowledgements}
G.W. thanks Dr. Kai Leong Chong, Dr. Luoqin Liu and Dr. Qi Wang for fruitful discussions. G.W. and R.J.A.M.S. acknowledge the financial support from ERC (the European Research Council) Starting Grant No. 804283 UltimateRB. This work was sponsored by NWO Science for the use of supercomputer facilities. We also acknowledge the Irene at Tr{\`e}s Grand Centre de Calcul du CEA (TGCC) under PRACE project 2019215098.
%We also acknowledge the national e-infrastructure of SURFsara, a subsidiary of SURF cooperation, the collaborative ICT organization for Dutch education and research
\FloatBarrier

\FloatBarrier
\section*{Declaration of interests}
The authors report no conflict of interest.
\FloatBarrier

\appendix

%%%%%%%%%%%%%%%%%%%%%%%%%%%%%%%%%%%%%%%%%%%%%%%%%%%%%%%%%%%
\section{Buoyancy components of off-center gravity} \label{appB}
\begin{figure}
\centering
\includegraphics[width=7cm]{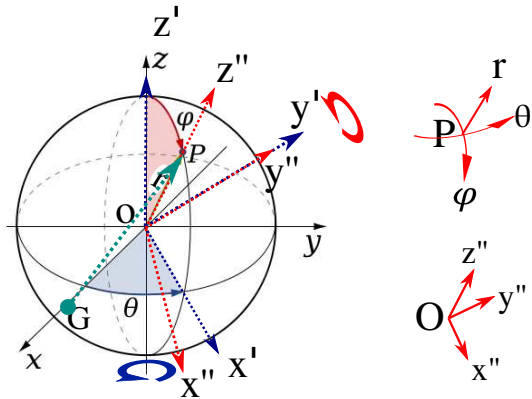}
 \caption{Schematic figure of the two-step transformation of a vector $\mathbf{GP}$ in Cartesian coordinate $xyz$ to spherical coordinate $\varphi \theta r$ ($x''y''z''$) frame. Circular arrows about $z$-axis and $y'$-axis show rotation axes.}
\label{fig:RotationMatrix}
\end{figure}

%\section{Transformation of gravity center to any random point} \label{appB}
%{\color{green}Buoyancy components of off-center gravity}}
%
The two-step transformation of a vector $\mathbf{GP}$ in Cartesian coordinate $xyz$ to spherical coordinate $\varphi \theta r$ ($x''y''z''$) is shown in figure \ref{fig:RotationMatrix}. A random fixed point $G(x_G,y_G,z_G)$ is representing the gravity center location. The other random point $P$ with $P(r\sin{\varphi} \cos{\theta},r\sin{\varphi} \sin{\theta}, r\cos{\varphi})$ is a moving point representing a fluid parcel. Initially, both $P$ and $G$ are in Cartesian coordinates $xyz$. Using two steps we will transfer $\mathbf{GP}$ to spherical coordinates $\varphi \theta r$ ($x''y''z''$).

\noindent (1) Rotating $xyz$ to $x'y'z'$ about $z$ by an angle $\theta$ using the right-hand side rule. The corresponding rotation matrix is 
\begin{equation}
R_{z}(\theta)=\left[\begin{array}{ccc}
{\cos \theta} & {-\sin \theta} & {0}.\\
{\sin \theta} & {\cos \theta} & {0} \\
{0} & {0} & {1}
\end{array}\right]
\end{equation}
(2) Rotating $x'y'z'$ to $x''y''z''$ about $y'$ by an angle $\varphi$ using the right-hand side rule. The corresponding rotation matrix is
\begin{equation}
R_{y}(\varphi)=\left[\begin{array}{ccc}
{\cos \varphi} & {0} & {\sin \varphi} \\
{0} & {1} & {0} \\
{-\sin \varphi} & {0} & {\cos \varphi}
\end{array}\right]
\end{equation}
After the above two steps we obtain that $x''y''z''$ overlaps with $\varphi \theta r$. The transformation matrix $\mathcal{M}_{\theta \varphi}$ from $\mathbf{GP}_{xyz}$ to $\mathbf{GP}_{x''y''z''}$ is defined by
\begin{equation}
\begin{aligned}
\mathcal{M}_{\theta \varphi}=R_{z}(\theta) \times R_{y}(\varphi) &=
\left[\begin{array}{ccc}
{\cos \theta} & {-\sin \theta} & {0} \\
{\sin \theta} & {\cos \theta} & {0} \\
{0} & {0} & {1}
\end{array}\right]
\left[\begin{array}{ccc}
{\cos \varphi} & {0} & {\sin \varphi} \\
{0} & {1} & {0} \\
{-\sin \varphi} & {0} & {\cos \varphi}
\end{array}\right] \\
&=
\left[\begin{array}{ccc}
{\cos \varphi} {\cos \theta} & {-\sin \theta} & {\sin \varphi} {\cos \theta} \\
{\cos \varphi} {\sin \theta} & {\cos \theta} & {\sin \varphi} {\sin \theta} \\
{-\sin \varphi} & {0} & {\cos \varphi}
\end{array}\right]
\end{aligned}
\end{equation}
In the Cartesian coordinate frame $\mathbf{GP}_{xyz}=(r\sin{\varphi} \cos{\theta}-x_G,~ r\sin{\varphi} \sin{\theta}-y_G,~ r\cos{\varphi}-z_G)$. Using the above transformations we find that $\mathbf{GP}$ in spherical coordinates $x''y''z''$ is:
\begin{equation}
\begin{aligned}
\mathbf{GP}_{x''y''z''}=\mathbf{GP}_{xyz} \times \mathcal{M}_{\theta \varphi}=
(- x_G \cos\varphi \cos\theta-y_G \cos\varphi \sin\theta + z_G \sin\varphi,&\\
 x_G \sin\theta - y_G \cos\theta,&\\
 r - x_G \sin\varphi \cos\theta -y_G\sin\varphi \sin\theta -z_G\cos\varphi&)
\end{aligned}
\end{equation}
In the dimensionless equations for the velocity $\mathbf{u}$, the buoyancy term is given by $gT\mathbf{GP}/L_{GP}$, where
\begin{equation}
L_{GP}=|\mathbf{GP}|=\sqrt{(r \sin\varphi \cos\theta - x_G)^2 + (r\sin \varphi \sin \theta-y_G)^2+(r \cos\varphi-z_G)^2}
\end{equation}
Three components in $\hat{\varphi}$, $\hat{\theta}$ and $\hat{r}$ directions can be decomposed as follows
\begin{equation}
\begin{aligned}
  C_{g,\theta}&=(x_G \sin\theta - y_G \cos\theta)/L_{GP} \\
  C_{g,r}&=(r - x_G \sin\varphi \cos\theta -y_G\sin\varphi \sin\theta -z_G\cos\varphi)/L_{GP} \\
  C_{g,\varphi}&=(- x_G \cos\varphi \cos\theta-y_G \cos\varphi \sin\theta + z_G \sin\varphi)/L_{GP}
\end{aligned}
\label{eq:C_g}
\end{equation}

%%%%%%%%%%%%%%%%%%%%%%%%%%%%%%%%%%%%%%%%%%%%%%%%%%%%%%%%%%%
\section{Code validations}\label{appA}

We extensively validated our code against the results presented by \cite{Gastine2015}, who performed a systematic parameter study for spherical RB convection with $Pr=1$ covering radius ratios $0.2 \le \eta \le 0.95$ for different gravity profiles in table  \ref{tab:Table_2}. The table shows that the $Nu$ and volume averaged Reynolds number $Re_{rms}$ agree within $2\%$ with the results from \cite{Gastine2015} for all cases. To ensure that the off-center gravity is implemented correctly, we verified that the flow dynamics for the same $L_{\mathbf{OG}}$, see figure \ref{fig:sphere_coordinate}, are identical for five different gravity center locations. The last five lines of table \ref{tab:Table_2} confirm that both $Nu$ and $Re_{rms}$ obtained from these simulations are consistent.

\begin{table}
\centering
\begin{adjustbox}{width=1.0\textwidth}
\begin{tabular}{ccccccccc}
$Ra$     & $\eta$ &$n_g$ & $G(\theta,r,\varphi)$ & $N_\theta,N_r,N_\varphi$ & $Nu$ &$Nu$(ref) &$Re_{rms}$ &$Re_{rms}$(ref) \\
\hline
$7\times10^6$ & 0.3  &$1$  & $O$    & 181,73,181 & 6.42 & 6.40& 282.2 &287.2  \\
       & 0.3  &$0$  & $O$    & 181,73,181 & 8.19 & 8.15& 377.5 &377.8  \\
$3\times10^7$ & 0.3  &$1$  & $O$    & 251,93,163 & 9.39 & 9.38& 592.0 &595.5  \\
$5\times10^6$ & 0.35 &$1$  & $O$    & 257,55,257 & 6.80 & 6.74& 274.8 &274.1 \\
$3\times10^8$ & 0.35 &$1$  & $O$    & 577,119,577 & 21.47 & 21.23& 1815.6 &1824.8  \\
$5\times10^6$ & 0.6  &$0$  & $O$    & 257,71,257 & 11.85 & 11.68& 437.3 &442.7  \\
$1\times10^6$ & 0.6  &$-2$  & $O$    & 325,71,325 & 9.02 & 8.90& 255.4 &259.2  \\
$3\times10^4$ & 0.6  &$-2$  & $O$    & 129,49,97 & 3.42 & 3.40& 44.1 &44.0  \\
       &    &$-2$  & $(0,r_i/2,\pi/2)$  & 129,49,97 & 3.197 & --& 40.8&-- \\
       &    &$-2$  & $(\pi/2,r_i/2,\pi/2)$& 129,49,97 & 3.187 & --& 40.8&-- \\
       &    &$-2$  & $(0,r_i/2,0)$    & 129,49,97 & 3.191 & --& 41.1&-- \\
       &    &$-2$  & $(\pi/4,r_i/2,\pi/4)$& 129,49,97 & 3.182 & --& 40.9&-- \\
       &    &$-2$  &$(\pi/7,r_i/2,3\pi/5)$& 129,49,97 & 3.186 & --& 41.0&-- \\
       
\end{tabular}
\end{adjustbox}
 \caption{Parameters of the spherical RB simulations with $Pr=1$, which are compared to the results ($Nu$(ref) and $Re_{rms}$(ref)) of \cite{Gastine2015}. $G(\theta,r,\varphi)$ is the gravity center location. $N_{\theta,r,\varphi}$ indicates the number of grid points in the longitudinal, radial, and co-latitudinal direction, respectively. The last five simulations for $Ra=3 \times 10^4$ are used to verify that the coefficients $C_{g,i}$, see Eq.\ \eqref{eq:C_g}, are correctly implemented by putting the gravity center $G$ in five different locations with the same $L_{\mathbf{OG}}$, see figure \ref{fig:sphere_coordinate}.}
 \label{tab:Table_2} 
\end{table}

%%%%%%%%%%%%%%%%%%%%%%%%%%%%%
\section{Spectral analysis in spherical coordinates}\label{appC}
This appendix describes the spectral analysis procedure used in this study. For further reading on this topic, we refer to p.122-p.145 of \cite{macrobert1947spherical}. In planar configurations, Fourier transformations can be applied when performing spectral analysis. However, for a spherical configuration, the basis wave functions are found by solving the Laplace equation. Analogous to the complex exponential in planar configuration, the spherical harmonics ${Y}_{\ell}^{m}({\theta}, {\varphi})$ read
\begin{equation}
{Y}_{\ell}^{m}({\theta}, {\varphi})={P}_{\ell}^{m}({\varphi}) e^{i m {\theta}}
\end{equation}
where $\ell$ and $m$ ($-\ell \le m \le \ell$) represent the wavenumber along a meridian and the equatorial plane, respectively. The polar angle $\varphi$ ranges from $0$ to $\pi$, and $\theta$ is the azimuthal angle, which ranges from $0 \leq \theta \leq 2\pi$. $P_{\ell}^{m} (\varphi)$ are the associated Legendre functions. For computational purposes the normalized associated Legendre function \eqref{eq:norm_2} are more attractive since the $P_{\ell}^{m}(\varphi)$ can overflow the computer \citep{swarztrauber1979spectral}.
\begin{equation}
\bar{P}_{\ell}^{m}(\varphi)=\left[\frac{2 \ell+1}{2} \frac{(\ell-m) !}{(\ell+m) !}\right]^{1 / 2} P_{\ell}^{m}(\varphi)
\end{equation}
The use of the spherical harmonics for approximating functions on a sphere is motivated by the fact that they form a complete system of orthogonal functions. Therefore, any function $f(\theta, \varphi)$, which is continuous and has continuous derivatives up to second-order, may be expanded in an absolutely and uniformly convergent series. 
\begin{equation}
f(\theta, \varphi)=\sum_{\ell=0}^{\infty} \sum_{m=-\ell}^{\ell} \bar{P}_{\ell}^{m}(\varphi)\left(a_{\ell, m} \cos (m \theta)+b_{\ell, m} \sin (m \theta)\right)
\label{eq:harmonics}
\end{equation}
where the expansion coefficients are given by 
\begin{equation}
a_{\ell, m}=\frac{1}{\pi} \int_{0}^{2 \pi} \int_{0}^{\pi} f(\theta, \varphi) \bar{P}_{\ell}^{m}(\varphi) \cos (m \theta) \sin \varphi d \theta d \varphi
\label{eq:alm}
\end{equation}
\begin{equation}
b_{\ell, m}=\frac{1}{\pi} \int_{0}^{2 \pi} \int_{0}^{\pi} f(\theta, \varphi) \bar{P}_{\ell}^{m}(\varphi) \sin (m \theta) \sin \varphi d \theta d \varphi
\label{eq:blm}
\end{equation}
The spherical harmonics ${Y}_{\ell}^{m}({\theta}, {\varphi})$ are orthonormal,
\begin{equation}
\frac{1}{2\pi} \int_{\varphi=0}^{\pi} \int_{\theta=0}^{2 \pi} Y_{\ell}^{m} Y_{\ell^{\prime}}^{m^{\prime} *} d \Omega=\delta_{\ell \ell^{\prime}} \delta_{m m^{\prime}},
\end{equation}
where $\delta_{i,j}$ is the Kronecker delta and $d \Omega=\sin \varphi d \theta d \varphi$ is the differential solid angle in spherical coordinates. Coefficients (\ref{eq:alm}) and (\ref{eq:blm}) are obtained by solving (\ref{eq:harmonics}) on a uniform lateral grid by using the SPHEREPACK library \citep{adams1999spherepack}.

Here we define the power spectrum on a lateral surface as
\begin{equation}
\begin{aligned}
&C_{\ell, m}=a^2_{\ell, m}+b^2_{\ell, m}\\
&C_{m}=\frac{1}{n-m+1} \sum_{\ell=m}^{n} C_{\ell, m},~n=\mbox{min}(N_\varphi-1,(N_\theta+1)/2)\\
\end{aligned}
\label{eq:Clm_Cl_Cm}
\end{equation}
where $C_{\ell, m}$ represents the power spectrum in lateral directions, and $C_{m}$ is the power spectrum in the longitudinal direction. Radial integration and temporal averaging are used to obtain $\langle C_{m} \rangle$.

\FloatBarrier
\bibliographystyle{jfm}
\bibliography{jfm-instructions}

\end{document}